\documentclass[a4paper]{jpconf}
\usepackage{graphicx}
\usepackage{units}
\usepackage{amssymb}
\usepackage{url}
\begin{document}
\title{Understanding time-resolved images of AWAKE proton bunches}

\author{Marlene Turner$^{1}$ and Patric Muggli$^{2}$}
\address{$^1$ CERN, Geneva, Switzerland}
\address{$^2$ Max Planck Institute for Physics, Munich, Germany}

\ead{marlene.turner@cern.ch}

\begin{abstract}
This article details how images of proton microbunch trains obtained from streak camera measurements may differ from actual microbunch trains inside the plasma, at the plasma exit. %
We use the same procedure as when comparing simulation results with measurements: create a particle distribution at the plasma exit using particle-in-cell simulations, propagate it to the location of the measurement and add diagnostic apertures and instrument resolution. %
From comparing distributions, we identify that changes in microbunch divergence and/or dimensions along trains result in differences between the charge distribution in reality and in the measurement. %
Additionally, we observe that instrument resolution reduces the observed modulation depth, with more reduction for shorter microbunches. %

\end{abstract}

\section{Introduction}

Plasma wakefield acceleration~\cite{bib:tajima,bib:chen} is a novel concept for particle acceleration, offering access to GV/m accelerating gradients. %
The concept works as follows: drivers (bunches or laser pulses driving wakefield) excite wakefields in plasma; witness bunches take energy from wakefields and accelerate. %
Intense laser pulses or dense particle bunches are ideal wakefield drivers when their length is on the order of the plasma wavelength $\sigma_z \sim \lambda_{pe}/\sqrt{2} = \sqrt{2}\pi c/\omega_{pe}$ ($\omega_{pe}=\sqrt{\frac{n_{pe} e^2}{\epsilon_0 m_e}}$, with $c$ the vacuum speed of light, $n_{pe}$ the plasma electron density, $e$ the electron charge, $\epsilon_0$ the vacuum dielectric constant and $m_e$ the electron mass, typically $\mu$m to mm) and their transverse size is on the order of the plasma skin depth $\sigma_r \sim c/\omega_{pe}$. %

Accelerating particles to very high energies ($\sim$TeV) in single plasma stages requires energetic drivers. %
Sufficiently energetic drivers exist, for example, proton bunches at CERN (containing tens to hundreds of kilo-Joules (kJ) of energy and a momentum of \unit[400 to 7000]{GeV/c} per particle) or pulses from disk lasers ($\sim$1 kJ of energy). %
However, these have lengths corresponding to many plasma wavelengths (e.g. \unit[$\sigma_z\sim6$]{cm} for CERN proton bunches), at plasma densities required to drive GV/m fields (\unit[$n_{pe}>1\times10^{14}$]{cm$^{-3}$}). %
They only drive wakefields effectively after their density is modulated, e.g. using the self-modulation process. %
Using energetic and self-modulated (proton bunch) drivers to accelerate particles to very high energies in a single plasma stage is the goal of the AWAKE experiment at CERN. %

The self-modulation (SM) process occurs when drivers longer than $\lambda_{pe}$ propagate in plasma. %
During the process, an initially uniform bunch distribution is transformed into a train of microbunches~\cite{bib:kumar} and wakefields grow both along the plasma and along the bunch until saturation~\cite{bib:marlenePRL,bib:marlenePRAB}. %
After saturation, microbunches are separated by the plasma wavelength ($\lambda_{pe}$), have a bunch length shorter than $\lambda_{pe}$ and resonantly drive wakefields to large amplitudes. 

Each microbunch finds its transverse equilibrium with the wakefields. %
Microbunches can thus propagate with constant radii over long plasma distances~\cite{bib:lotovradius}. %
However, after the plasma exit, they diverge according to their initial radius, emittance, and energy. %
Therefore their transverse size is different than in the plasma, when measured downstream of the plasma exit. %

\begin{figure}
    \centering
    \includegraphics[width=0.7\columnwidth]{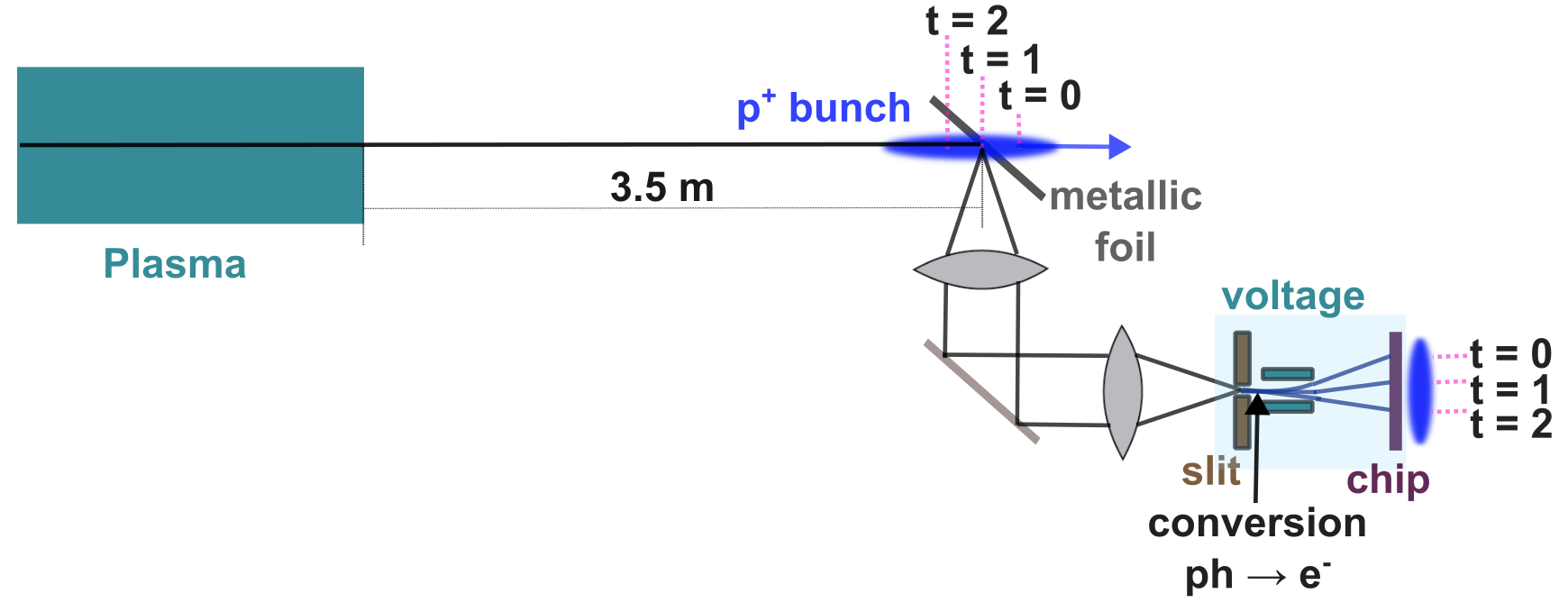}
    \caption{Schematic drawing of the streak camera measurement setup. %
    OTR light emitted at the metallic foil (screen) is imaged onto the entrance slit of a streak camera. %
    Inside the camera, the photons are converted to electrons with the same temporal and spatial structure. %
    Electrons are accelerated and deflected by a transverse, time-varying electric field to obtain time resolved images (t=0, 1, 2) of thin slices of the proton bunch density distribution on a CCD. %
    }
    \label{fig:sc}
\end{figure}

We briefly describe the experimental measurement setup used to record images of bunch distributions (microbunch trains) in AWAKE~\cite{bib:karlRSI}. %
An aluminum-coated metallic foil (screen) is inserted into the proton beam path. %
When protons traverse the screen, optical transition radiation (OTR) is emitted.
The radiation is imaged onto the entrance slit of a streak camera (see Fig.~\ref{fig:sc}). %
The slit selects an approximately \unit[80]{$\mu$m}-wide slice of the light/bunch around its axis. %
The camera then produces a time-resolved image of the bunch charge density distribution $n_b(x,t)$~\cite{bib:karl,bib:marlenePRL,bib:falk}. 
In this manuscript, we convert the time-axis to a spatial-axis $n_b(x,\xi)$, where $\xi = ct$ is the coordinate along the bunch. %

Since the screen is located \unit[3.5]{m} downstream of the plasma exit, the bunch distribution recorded at the screen is different from the one at the plasma exit. %
This difference arises due to the evolution that occurs over the \unit[3.5]{m} of vacuum propagation downstream of the plasma exit. %
Protons carry large (mostly longitudinal) momentum ($p_z\approx p$=\unit[400]{GeV/c}). %
Their longitudinal momentum change due to wakefields along the plasma is less than \unit[10]{GeV/c} (which corresponds to $<$\unit[1]{GV/m} amplitude of longitudinal wakefields over \unit[10]{m}). %
Therefore, velocity changes and dephasing between protons are negligible and do not cause microbunch divergence changes. %

However, the transverse momentum of protons in the bunch is much smaller ($p_{\perp}=\frac{\epsilon_g}{\sigma_{r0}} p$ = \unit[$\sim$400]{MeV/c}, where $\epsilon_g$ is the geometric emittance and $\sigma_{r0}$ the transverse rms bunch size at its waist) than the longitudinal one. %
Interaction with transverse wakefields, typically increasing along the bunch due to resonant wakefield excitation, increases the transverse proton momentum significantly~\cite{bib:marlenePRL}. %
Protons (and microbunches) therefore diverge according to their transverse momentum distribution during vacuum propagation. %
In addition, the resolution (both temporal and spatial) of the streak camera measurement is limited. %

In this article, we use numerical simulations to compare a bunch distribution downstream of the plasma exit differs with the one at the plasma exit and describe changes. %
This is important to understand how AWAKE streak camera measurements of microbunch trains are different than the microbunch trains at the plasma exit, which drive wakefields in the plasma. %

\section{Simulation of bunch distribution}

To study bunch propagation downstream of the plasma exit, we simulate the self-modulation of a long bunch in plasma using LCODE~\cite{bib:LCODE}. %
LCODE is a 2D3v-cylindrical, quasi-static, particle-in-cell code based on a fluid model. %
For illustration, we use the following beam (CERN-SPS-like proton bunches) and plasma parameters~\cite{bib:AWAKE} as simulation input: waist radius \unit[$\sigma_{r_0}=160$]{$\mu$m}, length \unit[$\sigma_{\xi}=6$]{cm}, population \unit[$N_b=3\times10^{11}$]{protons} and normalized emittance \unit[$\epsilon_N=2.6$]{mm mrad}, plasma length \unit[$L_p=10$]{m}, plasma density \unit[$n_{pe}=7\times10^{14}$]{cm$^{-3}$}, uniform with a \unit[3]{\%} density step at \unit[1.25]{m}~\cite{bib:lotovstep} and plasma radius \unit[$r_p=1$]{mm}. %
The SM seed (sharp, rising proton bunch density edge) is placed at $\xi=0$ (which is \unit[200]{ps} or \unit[60]{mm} ahead of the bunch center and located at \unit[$\xi=60$]{mm} on the Figures). %
These parameters lead to a particularly long microbunch train at the plasma exit. 

From the simulation result, we then create an image, similar to a streak camera measurement. %
We project the 2D3v (r, $\xi$, $p_r$, $p_{\phi}$) parameters of macro-particles into 3D ($x,y,\xi$) geometry, where $x$ is the horizontal, $y$ is the vertical coordinate, and $\xi$ is the co-moving coordinate along the bunch. %
We calculate particle positions ($x,y$) by projecting their radial positions ($r$) onto the horizontal and vertical axes, assigning random (uniformly distributed) angles $\theta$ from \unit[0 to 2$\pi$]{rad} to macro-particles ($x=r\cdot \textrm{cos}(\theta)$, $y=r\cdot \textrm{sin}(\theta)$). %

\begin{figure}[htb!]
    \centering
    \includegraphics[width=0.7\columnwidth]{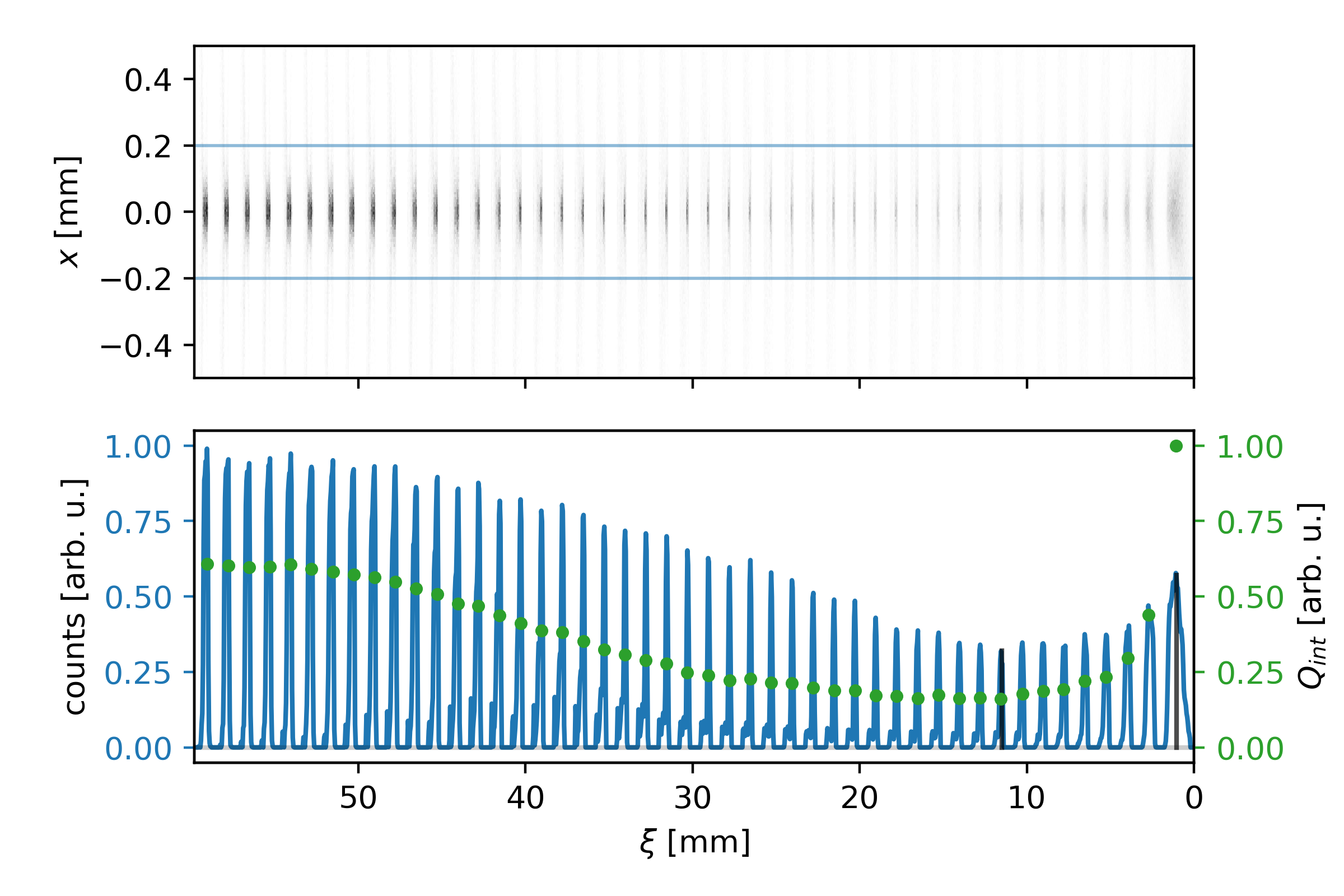}
    \caption{Top: 2D histogram of the proton bunch distribution at the plasma exit after self-modulation, including all protons in plasma. %
    Blue horizontal lines indicate $\pm c/\omega_{pe}$. %
    Bottom: Vertical sum of protons contained within $c/\omega_{pe}$ (blue solid line) and within the plasma (blue transparent dashed line) as well as the charge per microbunch $Q_{int}$ (green symbols) within $\pm c/\omega_{pe}$ and $\pm \lambda_{pe}/2$ from the microbunch peak. %
    The horizontal gray line indicates the minimum signal level of the blue line. %
    Vertical gray lines mark the modulation depth of microbunch Nr. 1 and 9. %
    The variable $\xi$ indicates the position along the bunch and $\xi=0$ is the location of the seed. %
    The bunch propagates to the right.
    }
    \label{fig:plasmaexit}
\end{figure}

On the top panel of Fig.~\ref{fig:plasmaexit}, we present the simulated proton density distribution along the bunch at the location of the plasma exit, for all protons residing inside of the plasma. %
The SM process has saturated~\cite{bib:marlenePRAB}. %
Microbunches formed in the focusing wakefield phases are spaced by the plasma wavelength (\unit[$\lambda_{pe}=1.2$]{mm} for this $n_{pe}$) and are clearly visible. %

The blue semi-transparent dashed line shows the vertical sum of all protons in plasma. %
All protons within approximately one plasma skin depth (indicated by the blue horizontal lines $x<c/\omega_{pe}$=\unit[200]{$\mu$m} on the top panel Fig.~\ref{fig:plasmaexit}) can effectively contribute to driving wakefields. %
Therefore, protons with radial positions smaller than $c/\omega_{pe}$ are shown in a vertical sum (blue line) on the bottom panel of Fig.~\ref{fig:plasmaexit}. %
The (relative) number of protons reaches zero between microbunches, as expected in a fully modulated bunch train. %

We calculate the charge per microbunch $Q_{int}$ (green symbols) by integrating counts over $\pm c/\omega_{pe}$ and over $\lambda_{pe}$ ($\pm \lambda_{pe}/2$ from the peak). %
The value of $Q_{int}$ approximately triples from $\xi\approx$ \unit[10]{mm} to \unit[60]{mm} due to the increasing charge along the initially Gaussian proton bunch. %
The even larger $Q_{int}$ between $\xi\approx$ \unit[0 and 10]{mm} is caused by the first few microbunches being longer than the following ones. %

\section{Propagation to the screen}
To understand the difference between the bunch distribution at the plasma exit, i.e., that of bunch driving wakefields, and the measured distributions using the streak camera setup, we perform the following steps: %

\begin{enumerate}
    \item Calculate the transverse positions of the protons (macro-particles) at the streak camera screen as $x_s,y_s=x,y+\frac{p_{x,y}}{\gamma_p m_p}\frac{\Delta z}{c}$, where $\gamma_p$ is the relativistic factor, $m_p$ the mass of the proton, and $\Delta z$=\unit[3.5]{m} the distance between the exit of the plasma and the screen. %
    We obtain horizontal and vertical particle momenta ($p_x,p_y$) by projecting the proton transverse momentum ($p_{\perp} = \sqrt{p_r^2+p_{\phi}^2}$) using $\theta$ as well as $\phi$ = arctan$(p_{\phi}/p_r)$. %
    \item Reduce the distribution to only include protons with vertical positions smaller than \unit[$\pm40$]{$\mu$m} from the axis to select for light signal captured by the streak cameras slit. %
    \item Include the streak camera instrument resolution of \unit[$\sigma_{\Delta x}$=$180$]{$\mu$m} in the transverse dimension and \unit[0.6]{\%} times the window length ($\sigma_{\Delta\xi}$= (\unit[200]{ps} $\times$ \unit[0.6]{\%})=\unit[1.2]{ps} $\mathrel{\widehat{=}}$ \unit[0.36]{mm}) temporally. %
    We include these by applying corresponding Gaussian convolutions (blurring)~\cite{bib:blur} to the 2D images. %
\end{enumerate}

\begin{figure}[htb!]
    \centering
    \includegraphics[width=0.7\columnwidth]{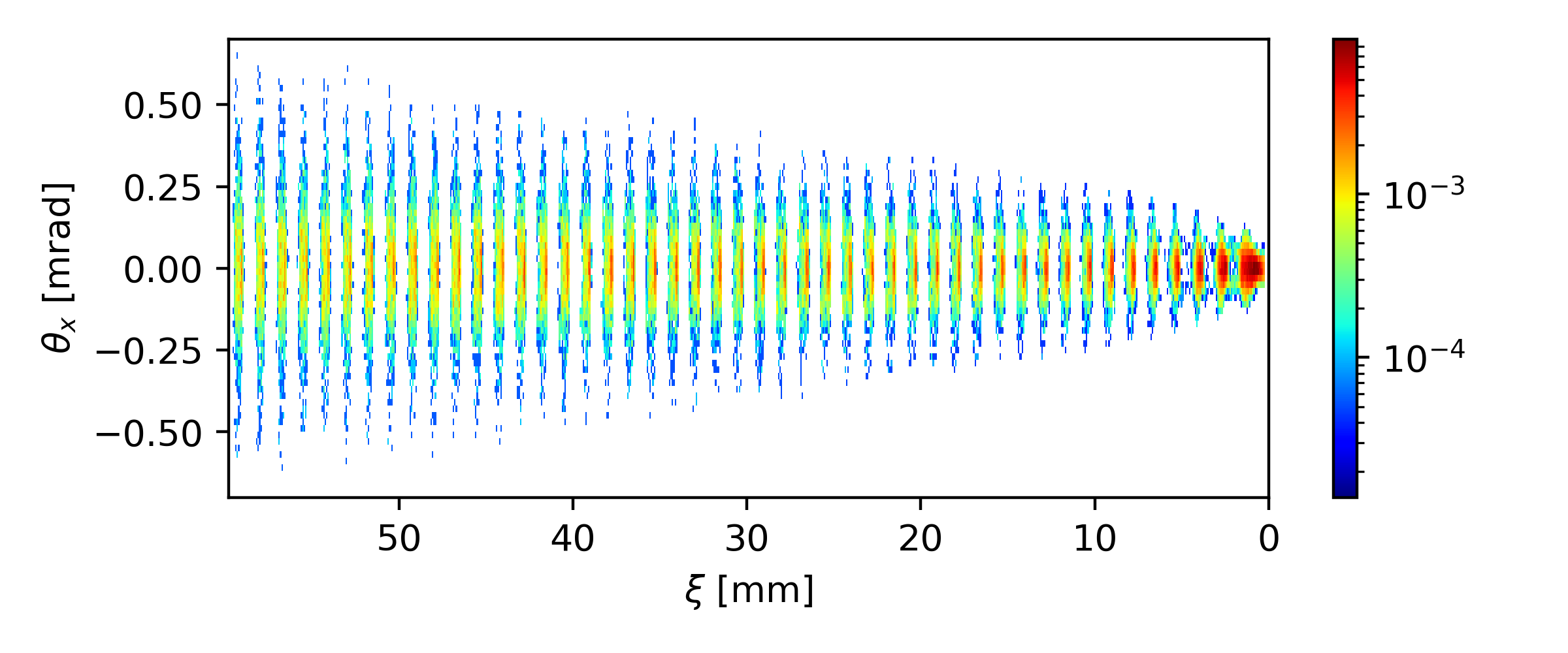}
    \caption{Transverse momentum ($\theta_{x}=p_x/p_{\parallel}$) distribution of protons arriving within the streak camera slit. %
    The logarithmic color scale was chosen to enhance the visibility of the particles with \unit[$|\theta_x|>0.25$]{mrad}. %
    }
    \label{fig:pperp} 
\end{figure}

On Fig.~\ref{fig:pperp}, we display the divergence distribution of protons at the plasma exit as the angle $\theta_{x}=p_x/p_{\parallel}$. %
A divergence of \unit[0.1]{mrad} over \unit[3.5]{m} leads to a transverse size increase of \unit[350]{$\mu$m}, i.e., an increase comparable to the unmodulated transverse bunch size at the location of the plasma exit (\unit[$\sim540$]{$\mu$m}). %
The increase along $\xi$ (both in the core (0$<\theta_x<\sim0.1$) and the wings ($\theta_x>\sim0.1$)) results either from microbunches experiencing stronger (focusing) fields or emittance growth during propagation in plasma. %

\begin{figure}[htb!]
    \centering
    \includegraphics[width=0.7\columnwidth]{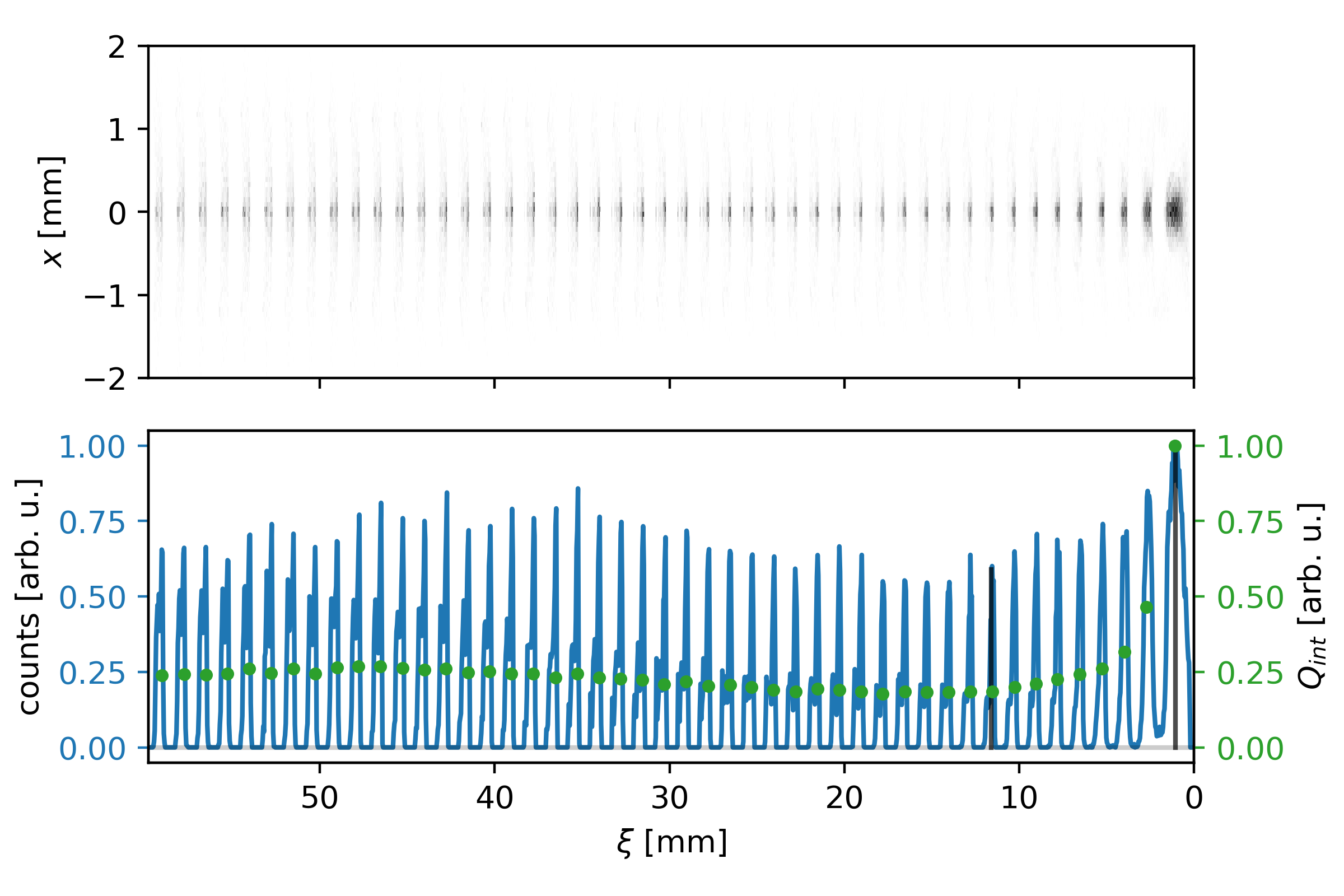}
    \caption{Top: Proton bunch distribution (shown on the top of Fig.~\ref{fig:plasmaexit}), propagated over \unit[3.5]{m} of vacuum to the location of the streak camera screen. %
    Additionally, a vertical cut of \unit[$y<\pm$40]{$\mu$m} (corresponding to the streak camera slit) is applied. %
    Bottom: Vertical sum of the distribution (blue line) shown on the top panel as well as the charge per microbunch $Q_{int}$ (green symbols) within $\pm \lambda_{pe}/2$ from their peak. %
    The horizontal gray line indicates the minimum signal level of the blue line. %
    Vertical gray lines mark the modulation depth of microbunch Nr. 1 and 9. %
    }
    \label{fig:scnoblur}
\end{figure}

On Fig.~\ref{fig:scnoblur}, we present the proton bunch density distribution (top) and vertical projection (bottom) after vacuum propagation (i) and slit implementation (ii), but without the blurring that imitates the limited instrument resolution. %
Due to divergence during vacuum propagation the transverse microbunch extent is larger (\unit[FWHM=500-800]{$\mu$m}) than at the plasma exit (\unit[FWHM=150-200]{$\mu$m}, on Fig.~\ref{fig:plasmaexit}). %

Comparing vertical sums (blue lines) and charges per microbunch $Q_{int}$ (green symbols) on the bottom panels of Figs.~\ref{fig:plasmaexit} and~\ref{fig:scnoblur} illustrates how the observed proton bunch density distribution changes. %
At the plasma exit (Fig.~\ref{fig:plasmaexit}) $Q_{int}$ (green symbols) decreases by \unit[84]{\%} (from 1 to 0.16 between \unit[$\xi=$0-10]{mm}) and then increases by \unit[44]{\%} (to 0.6) for \unit[$\xi>$10]{mm}. %
However, at the screen (Fig.~\ref{fig:scnoblur}), $Q_{int}$ (green symbols) decreases by \unit[79]{\%} (from 1 at \unit[$\xi=$1]{mm} to 0.21 at \unit[$\xi=11.5$]{mm}) and is approximately constant for \unit[$\xi>$10]{mm}. %
Because the microbunch divergence is increasing along $\xi$ (see Fig.~\ref{fig:pperp}) $Q_{int}$ remains approximately flat (rather than increase as on Fig.~\ref{fig:plasmaexit}) for \unit[$\xi>$10]{mm} on Fig.~\ref{fig:scnoblur}. %

\begin{figure}[htb!]
    \centering
    \includegraphics[width=0.7\columnwidth]{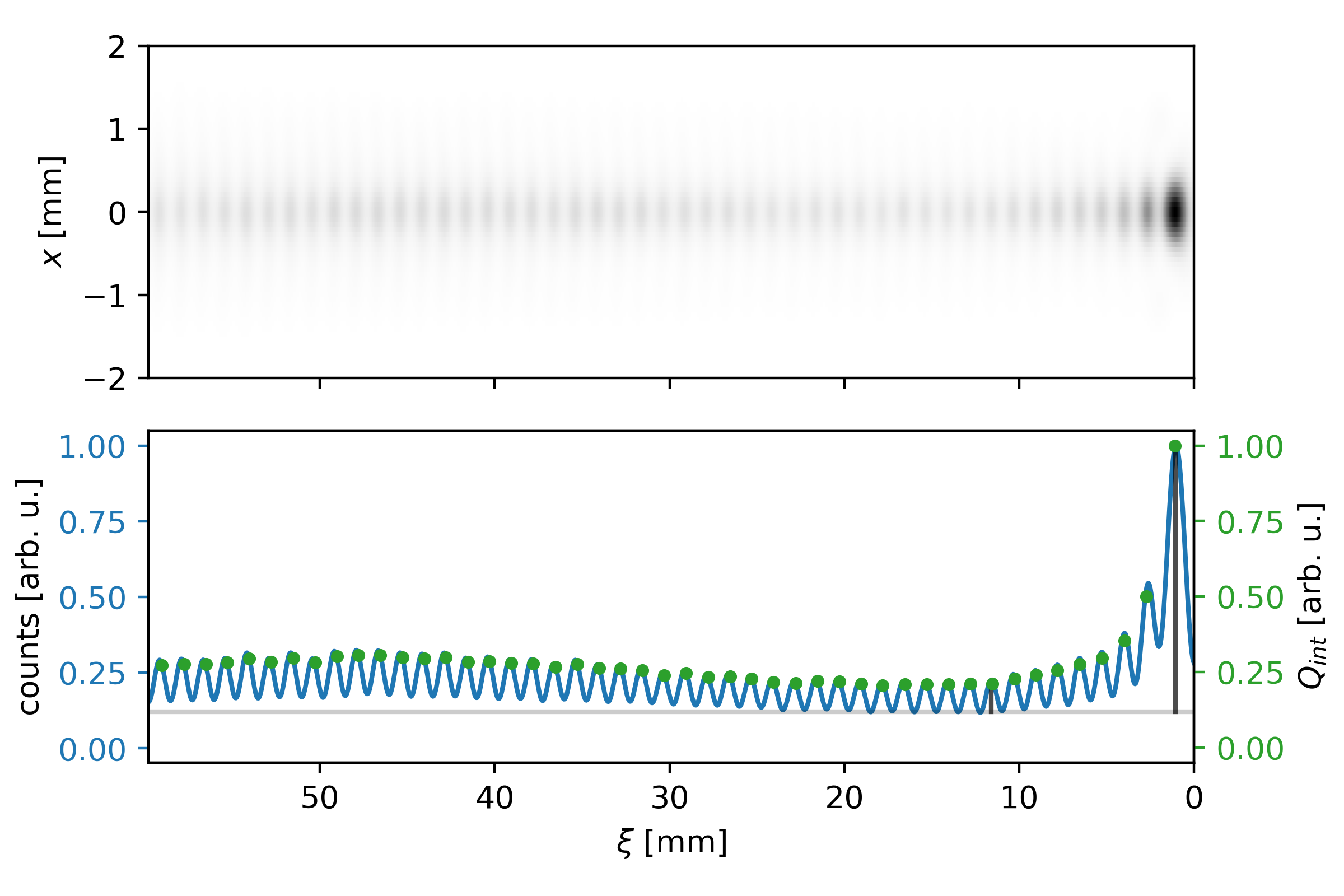}
    \caption{Top: Simulated proton bunch distribution, propagated over \unit[3.5]{m} of vacuum to the location of the streak camera screen. %
    Additionally, a vertical cut of \unit[$y<\pm$80]{$\mu$m} (corresponding to the streak camera slit) and Gaussian blur (of \unit[360]{$\mu$m} in the temporal ($\xi$) and of \unit[180]{$\mu$m} in the spatial ($x$) dimensions) are applied. %
    Note that the vertical scale is different from that of Fig.~\ref{fig:plasmaexit}. %
    Bottom: Vertical sum of the distribution shown on the top panel (blue line). Green symbols indicate $Q_{int}$, which is the charge per microbunch. %
    The horizontal gray line indicates the minimum signal level of the blue line. %
    Vertical gray lines mark the modulation depth of microbunch Nr. 1 and 9. %
    }
    \label{fig:streakcamera}
\end{figure}

Next, we imitate the effect of instrument resolution by adding Gaussian blurring (iii) to the image displayed on Fig.~\ref{fig:scnoblur}. %
The resulting image i.e., the simulated streak camera image, is displayed on the top panel of Fig.~\ref{fig:streakcamera} and the corresponding vertical sum is shown on the bottom panel (blue line). %

We observe that blurring causes a further microbunch amplitude drop between \unit[$\xi$=1-12]{mm} (compare blue lines on the bottom of Figs.~\ref{fig:streakcamera} and~\ref{fig:scnoblur}), because of the changes in microbunch lengths $\sigma_{\xi}$. %
The first few microbunches are longer~\cite{bib:thesisannamaria} (i.e. \unit[$\sigma_{\xi}=0.4, 0.2$ and 0.15]{mm} for \unit[$\xi$=1, 2.5 and 3.95]{mm} compared to \unit[$\sigma_{\xi}\sim0.12\pm0.02$]{mm} for others in the window). %
Lengths are obtained from Gaussian fits to individual microbunches in the vertical sum (blue line, bottom panel of Fig.~\ref{fig:plasmaexit}). %
The absolute values of $\sigma_{\xi}$ (\unit[$\sigma_{\xi}=$0.12-0.4]{mm}) as well as changes of $\sigma_{\xi}$ (\unit[$\Delta\sigma_{\xi}$=0.28]{mm}) along $\xi$ are on the same order as the temporal instrument resolution \unit[$\sigma_{\Delta\xi}=0.36$]{mm}. %
Blurring by the same $\sigma_{\Delta\xi}$ affects shorter microbunches more as their observed length on the streak camera ($\sigma_{o,\xi}$) is determined by their original length combined in quadrature with the temporal instrument resolution ($\sigma_{o,\xi}=\sqrt{\sigma_{\xi}^2+\sigma_{\Delta\xi}^2}$). %
For example, the length of the microbunch located at \unit[$\xi=1$]{mm} increases from \unit[$\sigma_{\xi,1}=$0.4]{mm} to \unit[$\sigma_{o,\xi,1}=$0.53]{mm} (\unit[$\approx$30]{\%}). %
Bunches located at larger $\xi$ increase their length from \unit[$\sigma_{\xi}\sim$0.12]{mm} to \unit[$\sigma_{o,\xi}=$0.38]{mm} (\unit[$>300$]{\%}). %

Then, blurring further reduces microbunch modulation depth (compare blue lines in Figs.~\ref{fig:scnoblur} and~\ref{fig:streakcamera}).
This is because the signals from charges from one microbunch overlap with the ones from neighboring microbunches, due to temporal instrument resolution.
This effect is visible from the minimum of \unit[0.12]{counts} of the blue line on the bottom panel of Fig.~\ref{fig:streakcamera} compared to \unit[0]{counts} on the bottom panels of Figs.~\ref{fig:streakcamera} and~\ref{fig:scnoblur}. %
On Fig.~\ref{fig:scnoblur}, modulation depth decreases by \unit[41]{\%} from microbunch Nr.~1 (\unit[$\xi=1$]{mm}) to Nr.~9 (\unit[$\xi=11.5$]{mm}).
On Fig.~\ref{fig:streakcamera}, it decreases by \unit[89]{\%} over the same range. %

However, blurring leaves $Q_{int}$ mostly unchanged, as seen by comparing $Q_{int}$ on the bottom panels of Figs.~\ref{fig:scnoblur} and~\ref{fig:streakcamera}. %
This is because signals belonging to a microbunch are for the most part still contained within the integration range (\unit[$x<\pm2$]{mm} and \unit[$\xi<\pm\lambda_{pe}/2$]{mm} from the microbunch peak). %

\section{Discussion}
Depending on the beam and plasma parameters, the difference between distributions of protons at the plasma exit and those at the screen where the distributions are measured, can be caused by the divergence of the protons and/or by the limited resolution the streak camera (blurring in space and time). %
The effect of blurring becomes increasingly important with higher $n_{pe}$, and is significant when $\lambda_{pe}/4\lesssim\sigma_{\Delta\xi}$, as $\sigma_{\xi}\sim\lambda_{pe}/4$. %
The effect of divergence increases with increasing $n_{pe}$ and $N_b$, as wakefield amplitudes and therefore $\theta_{x,y}$ are larger and changing more along the bunch train. %

In this article, we apply the same procedure that is routinely applied to compare simulation results with measurements in AWAKE. %
To still be able to estimate the charge per microbunch, previous work either: %
\begin{itemize}
    \item included the transverse microbunch size to correct for the effect of the slit. %
    This is challenging because determination of size is non trivial and signal levels may be below detection threshold. %
    However, this technique can work well, especially at low plasma density (for example, in Ref.~\cite{bib:thesisannamaria} at \unit[$n_{pe}=0.96\times10^{14}$]{cm$^{-3}$}).
    \item obtained good agreement between the measurements and propagated simulation results, to then infer the microbunch train properties from the simulation result at the plasma exit. %
\end{itemize}

\section{Summary and conclusions}

We discuss how streak camera images of proton microbunch trains differ from actual distributions inside of the plasma, at the plasma exit. %
When measured downstream the plasma, the transverse microbunch size is larger than at the plasma exit because microbunches diverge during vacuum propagation. %
Because wakefield amplitudes change along the bunch train, microbunch divergence also changes. %
Microbunches with larger divergence appear with lower microbunch charge density because of the effect of the streak camera slit. %
Additionally, the limited instrument resolution reduces the observed microbunch modulation depth. %
The reduction is stronger for shorter the microbunches. %
Therefore, while streak cameras images show the effect of the self-modulation process on the density distributions of long bunches clearly, the observed microbunch charge density is different from the one at the plasma exit. %

\end{document}